\RequirePackage{fix-cm} % Fix LaTeX2e bugs.
\documentclass[a4paper, twoside, reqno, dvips, 12pt]{amsart}
\usepackage{fixltx2e}   % Fix LaTeX2e bugs.

\usepackage{etex}

\usepackage[latin1]{inputenc}
\usepackage[T1]{fontenc}

\usepackage{esint}
\usepackage{dsfont}
\usepackage{xspace}
\usepackage{amsgen}
\usepackage{amsthm}
\usepackage{amssymb}
\usepackage{amsmath}
\usepackage{wasysym}
\usepackage{upgreek}
\usepackage{amsfonts}
\usepackage{stmaryrd}
\usepackage{scalerel}
\usepackage{mathtools}
\usepackage{stackengine}

\usepackage{relsize}
\usepackage[mathcal, mathscr]{euscript}

\usepackage{mathrsfs}
\DeclareMathAlphabet{\mathscrbf}{OMS}{mdugm}{b}{n}

\usepackage{a4wide}

\usepackage[dvipsnames, table]{xcolor}
\definecolor{bckg}{RGB}{20.8, 20.8, 20.8}
\definecolor{oneblue}{rgb}{0.0, 0.0, 0.85}
\definecolor{Lightblue}{RGB}{214, 214, 214}
\definecolor{bluepigment}{rgb}{0.2, 0.2, 0.6}
\definecolor{charcoal}{rgb}{0.21, 0.27, 0.31}
\definecolor{denimblue}{rgb}{0.08, 0.38, 0.74}
\definecolor{Lightgray}{rgb}{0.89, 0.89, 0.89}
\definecolor{darkgrey}{rgb}{0.273, 0.281, 0.30}
\definecolor{darkelectricblue}{rgb}{0.33, 0.41, 0.47}

\usepackage[sort&compress, comma, square, numbers]{natbib}

\usepackage{psfrag}
\usepackage{graphicx}
\usepackage{adjustbox}
\usepackage[tight]{subfigure}
\usepackage{morefloats}
\usepackage{indentfirst}

\usepackage[usenames, dvipsnames, pdf]{pstricks}
\usepackage{epsfig}
\usepackage{pst-grad} % For gradients
\usepackage{pst-plot} % For axes

\usepackage{acronym}
\usepackage{microtype}
\usepackage[labelsep=period,%
            labelfont={bf,sf,color=bluepigment},%
            justification=raggedright]{caption}

\usepackage[perpage, symbol]{footmisc}

\usepackage[colorlinks,
           urlcolor=oneblue,
           linkcolor=denimblue,
           citecolor=NavyBlue,
           bookmarksopen=false,
           pdfpagemode=UseNone,
           pagebackref]{hyperref}

\usepackage[explicit]{titlesec}

\titleformat{\section}[block]
  {\color{NavyBlue}\Large\sffamily\bfseries}
  {}
  {0.0em}
  {\colorbox{bckg!5}{\strut\parbox{\dimexpr\linewidth-2\fboxsep\relax}{\thesection. #1}}}
  [\vspace*{0.33em}]

\titleformat{name=\section,numberless}[block]
  {\color{NavyBlue}\Large\sffamily\bfseries}
  {}
  {0.0em}
  {\colorbox{bckg!5}{\strut\parbox{\dimexpr\linewidth-2\fboxsep\relax}{#1}}}
  [\vspace*{0.33em}]

\titleformat{\subsection}
  {\color{NavyBlue}\large\sffamily\bfseries}
  {}
  {0.0em}
  {\colorbox{bckg!5}{\parbox{\dimexpr\linewidth-2\fboxsep\relax}{\thesubsection. #1}}}
  [\vspace*{0.33em}]

\titleformat{name=\subsection,numberless}
  {\color{NavyBlue}\Large\sffamily\bfseries}
  {}
  {0em}
  {\colorbox{bckg!5}{\parbox{\dimexpr\linewidth-2\fboxsep\relax}{#1}}}
  [\vspace*{0.33em}]

\titleformat{\subsubsection}
  {\color{bluepigment}\sffamily\normalsize\bfseries}
  {\thesubsubsection}
  {0.5em}
  {#1}
  [\vspace*{0.33em}]

\titleformat{\paragraph}[runin]
  {\color{bluepigment}\sffamily\small\bfseries}
  {}
  {0em}
  {#1}

\titlespacing{\section}{0.0em}{1.5em plus 2pt minus 2pt}%
{1.0em plus 2pt minus 2pt}[0em]
\titlespacing{\subsection}{0.5em}{1.5em plus 2pt minus 2pt}%
{1.0em}[0em]
\titlespacing{\subsubsection}{0.5em}{1.5em plus 2pt minus 2pt}%
{1.0em plus 2pt minus 2pt}[0em]

\usepackage{titletoc}

\contentsmargin{0.5em}
\newlength{\tocsep} 
\titlecontents{section}[\tocsep]
  {\addvspace{10pt}\bfseries\sffamily}
  {\contentslabel[\thecontentslabel]{\tocsep}}
  {}
  {\ \titlerule*[0.75pc]{.}\ \thecontentspage}
  []
\titlecontents{subsection}[\tocsep]
  {\addvspace{8pt}\sffamily}
  {\contentslabel[\thecontentslabel]{\tocsep}}
  {}
  {\ \titlerule*[0.5pc]{.}\ \thecontentspage}
  []
\titlecontents*{subsubsection}[\tocsep]
  {\addvspace{2pt}\footnotesize\sffamily}
  {}
  {}
  {\ \titlerule*[0.35pc]{.}\ \thecontentspage}
  [\\*]

\makeatletter
\def\@setauthors{%
  \begingroup
  \def\thanks{\protect\thanks@warning}%
  \trivlist
  \centering\footnotesize \@topsep30\p@\relax
  \advance\@topsep by -\baselineskip
  \item\relax
  \author@andify\authors
  \def\\{\protect\linebreak}%
  \textsc{\normalsize\textcolor{darkelectricblue}{\authors}}%
  \ifx\@empty\contribs
  \else
    ,\penalty-3 \space \@setcontribs
    \@closetoccontribs
  \fi
  \endtrivlist
  \endgroup
}
\def\@settitle{\begin{center}%
  \baselineskip14\p@\relax
    \bfseries
    \textsc{\Large\textcolor{charcoal}{\@title}}
  \end{center}%
}
\makeatother

\usepackage{enumitem}
\setlist[description]{%
  topsep=30pt,               % space before start / after end of list
  itemsep=5pt,               % space between items
  font={\bfseries\sffamily\color{NavyBlue}}, % if colour is needed
}

\usepackage{fancyhdr}
\usepackage{lastpage}

\newcommand*\Title{\textcolor{bluepigment}{Multi-symplectic structure of long internal waves}}
\newcommand*\Authors{\textcolor{bluepigment}{D.~Clamond \& D.~Dutykh}}
\newcommand*{\plogo}{\textcolor{gray}{{\texttt{arXiv.org} / \textsc{hal}}}} % Generic publisher logo

\numberwithin{equation}{section}

\newcommand{\ud}{\mathrm{d}}

\newcommand{\uD}{\mathrm{D}}

\newcommand{\ve}{\boldsymbol{e}}
\newcommand{\vu}{\boldsymbol{u}}
\newcommand{\vz}{\boldsymbol{z}}

\newcommand{\ie}{\emph{i.e.}\/ }
\newcommand{\eg}{\emph{e.g.}\/ }

\newcommand{\scal}{\boldsymbol{\cdot}}
\newcommand{\grad}{\boldsymbol{\nabla}}

\newcommand{\eqdef}{\mathop{\stackrel{\,\mathrm{def}}{:=}\,}}

\newcommand{\half}{{\textstyle{1\over2}}}
\newcommand{\third}{{\textstyle{1\over3}}}

\newcommand{\sixth}{{\textstyle{1\over6}}}

\newcommand{\twothird}{{\textstyle{2\over3}}}

\usepackage{acronym}
\acrodef{bvp}[BVP]{Boundary Value Problem}
\acrodef{NSWE}{Nonlinear Shallow Water Equations}

\begin{document}

\title[\Title]{Multi-symplectic structure of fully-nonlinear weakly-dispersive internal gravity waves}

\author[D.~Clamond]{Didier Clamond$^*$}
\address{Laboratoire J. A. Dieudonn\'e, Universit\'e de Nice -- Sophia Antipolis, Parc Valrose, 06108 Nice cedex 2, France}
\email{diderc@unice.fr}
\urladdr{http://math.unice.fr/~didierc/}
\thanks{$^*$ Corresponding author}

\author[D.~Dutykh]{Denys Dutykh}
\address{LAMA, UMR 5127 CNRS, Universit\'e Savoie Mont Blanc, Campus Scientifique, 
73376 Le Bourget-du-Lac Cedex, France}
\email{Denys.Dutykh@univ-savoie.fr}
\urladdr{http://www.denys-dutykh.com/}


%%% ----------------------------------------------------------------------- %%%

\begin{titlepage}
\thispagestyle{empty} % Remove page numbering on this page
\noindent
{\Large Didier \textsc{Clamond}}\\
{\it\textcolor{gray}{Universit\'e de Nice -- Sophia Antipolis, France}}\\[0.02\textheight]
{\Large Denys \textsc{Dutykh}}\\
{\it\textcolor{gray}{CNRS--LAMA, Universit\'e Savoie Mont Blanc, France}}
\\[0.08\textheight]

\vspace*{1.1cm}

\colorbox{Lightblue}{
  \parbox[t]{1.0\textwidth}{
    \centering\huge\sc
    \vspace*{0.7cm}
    
    \textcolor{bluepigment}{Multi-symplectic structure of fully-nonlinear weakly-dispersive internal gravity waves}
    
    \vspace*{0.7cm}
  }
}

\vfill % Whitespace between the title block and the publisher

\raggedleft     % Right-align all text
{\large \plogo} % Publisher and logo
\end{titlepage}

%%% ----------------------------------------------------------------------- %%%

\newpage
\thispagestyle{empty} % Remove page numbering on this page
\par\vspace*{\fill}   % Whitespace until the bottom
\begin{flushright} % Right-align all text
{\textcolor{denimblue}{\textsc{Last modified:}} \today}
\end{flushright}

%%% ----------------------------------------------------------------------- %%%

\newpage
\maketitle
\thispagestyle{empty}

%%% ------------------------------------------------------------------------ %%%

\begin{abstract}

In this short communication we present the multi-symplectic structure for the two-layer \textsc{Serre}--\textsc{Green}--\textsc{Naghdi} equations describing the evolution of large amplitude internal gravity long waves. We consider only a two-layer stratification with rigid bottom and lid for simplicity, generalisations to several layers being straightforward. This multi-symplectic formulation allows the application of various multi-symplectic integrators (such as \textsc{Euler} or \textsc{Preissman} box schemes) that preserve exactly the multi-symplecticity at the discrete level. 

\bigskip
\noindent \textbf{\keywordsname:} internal waves, two-layer flow, multi-symplectic structure, long waves, Green--Naghdi equations \\


\end{abstract}

%%% ----------------------------------------------------------------------- %%%

\newpage
\tableofcontents
\thispagestyle{empty}

%%% ----------------------------------------------------------------------- %%%

\newpage
\section{Introduction}

The density stratification in oceans appears naturally as the result of the existence of a thermocline, differences in the salinity and other similar mechanisms \cite{Garrett1979}. The density stratification supports the so-called `internal waves'. These are ubiquitous in the ocean and, comparing to surface waves, internal waves may have a huge amplitude of the order of hundreds of meters \cite{Garrett1975}. These waves play an important role in ocean dynamics and they attract permanent attention of several scientific communities. Compared to surface gravity waves, the physics of internal waves is richer and their modelling leads to more complicated equations, in general. 

Lagrangian and Hamiltonian formalisms are tools of choice in theoretical Physics, in particular for studying nonlinear waves. Quite recently, the multi-symplectic formalism have been proposed as an attractive alternative. This formulation generalises the classical Hamiltonian structure to partial differential equations by treating space and time on the equal footing \cite{Bridges1997}. Multi-symplectic formulations are gaining popularity, both for mathematical investigations 
and numerical modelling \cite{Bridges2001, Moore2003a}.

Multi-symplectic formulations of various equations modelling surface waves can be found in the literature. However, to our knowledge, no such formulations have been proposed for internal waves. In this note, we show that the multi-symplectic structure of a homogeneous fluid can be easily extended to fluids stratified in several homogeneous layers. For the sake of simplicity, we focus on two-dimensional irrotational motions of internal waves propagating at the interface between two perfect fluids, bounded below by an impermeable horizontal bottom and bounded above by an impermeable rigid lid.

The present article should be considered as a further step in understanding the underlying mathematical structure of an important model of long internal waves --- the so-called two-layer \textsc{Serre} equations \cite{Serre1953}. \textsc{Serre}'s equations are a shallow water (long waves) approximation for large amplitude waves. This model is sometimes referred to as {\em weakly-dispersive fully-nonlinear} and was first derived by \textsc{Serre} for surface waves \cite{Serre1953}. Its generalisation for two layers internal waves was apparently first obtained by \textsc{Bart\'el\'emy} \cite{Barthelemy1989}, both with a rigid lid and with a free surface. This model was later re-derived using different approaches \cite{Barros2007a, Barros2007, Choi1999a}.

The Hamiltonian formulation for the classical \textsc{Serre} equations describing the surface waves can be found in \cite{Johnson2002}, for example. However, this structure is non-canonical and quite non-trivial. The Hamiltonian formulation for travelling wave solutions for two-layer \textsc{Serre} equations can be found in \cite{Barros2007a}. In the present study, we propose a multi-symplectic formulation of two-layer \textsc{Serre} equations with the rigid lid. This work is a direct continuation of \cite{Chhay2016} where the multi-symplectic structure was proposed for the original \textsc{Serre} equations.

The results obtained in this study can be used to propose new structure-preserving numerical schemes to simulate the dynamics of internal waves. Indeed, now it is straightforward to apply the \textsc{Euler}-box or the \textsc{Preissman}-box schemes \cite{Moore2003a} to two-layers \textsc{Serre} equations. The advantage of this approach is that such schemes preserve \emph{exactly} the multi-symplectic form at the discrete level \cite{Bridges1997, Bridges2001}. To our knowledge, this direction is essentially open. We are not aware of the existing structure-preserving numerical codes to simulate \emph{internal} waves. This situation may be explained by higher complexity of these models compared to, \eg surface waves.

The present manuscript is organised as follows. In Section~\ref{sec:model}, we present a simple variational derivation of the governing equations. Their multi-symplectic structure is provided in Section~\ref{sec:ms} and the implied conservation laws are given in Section~\ref{sec:cons}. The main conclusions and perspectives of this study are outlined in Section~\ref{sec:concl}.

%%% ----------------------------------------------------------------------- %%%

\section{Model derivation}
\label{sec:model}

\begin{figure}
  \centering
  \includegraphics[width=0.8\textwidth]{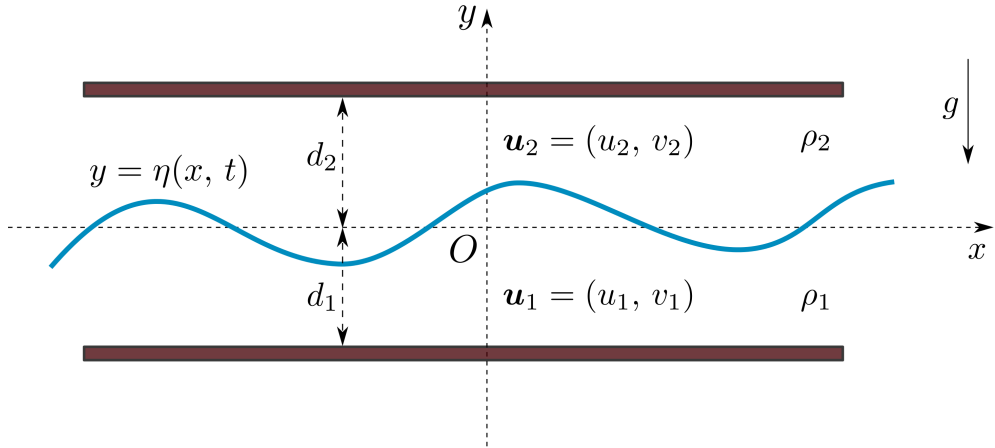}
  \caption{\small\em Sketch of the two-fluid domain.}
  \label{fig:sketch}
\end{figure}

We consider a two-dimensional irrotational flow of an incompressible fluid stratified in two homogeneous layers of densities $\rho_j$, subscripts $j=1$ and $j=2$ denoting the lower and upper layers, respectively. The fluid is bounded below by a horizontal impermeable bottom at $y=-d_1$ and above by a rigid lib at $y=d_2$, $y$ being the upward vertical coordinate such that $y=0$ and $y=\eta(x,t)$ are, respectively, the equations of the still interface and of the wavy interface (see Figure~\ref{fig:sketch}). Here, $x$ denotes the horizontal coordinate, $t$ is the time, $g$ is the downward (constant) acceleration due to gravity and surface tensions are neglected. The lower and upper thicknesses are, respectively, $h_1=d_1+\eta$ and $h_2 = d_2-\eta$, such that $h_1 + h_2 = d_1 + d_2 \eqdef D$ is a constant. Finally, we denote $\vu_j = (u_j, v_j)$ the velocity fields in the $j$-th layer ($u_j$ the horizontal velocities, $v_j$ the vertical ones). We shall now derive the fully-nonlinear weakly-dispersive long wave approximation \textsc{Serre}-like equations \cite{Barthelemy1989, Choi1999a} following a variational approach initiated in \cite{Staub1972} for the (one-layer) classical \textsc{Serre} equations. Further relations are given in the Appendix.

\subsection{Ansatz}

In order to model long waves in shallow water with rigid horizontal bottom and lid, the velocity fields in each layer is approximate as
\begin{gather}\label{defuvsej}
  u_j(x,y,t)\ \approx\ \bar{u}_j(x,t), \qquad v_j(x,y,t)\ \approx\ \left((-1)^jd_j-y\right)\bar{u}_{jx},
\end{gather}
where $\bar{u}_j$ is the horizontal velocity averaged over the $j$-th layer, i.e., $\bar{u}_1 \eqdef h_1^{\,-1}\int_{-d_1}^\eta u_1\/\ud y\,$, $\bar{u}_2\eqdef h_2^{\,-1}\int_\eta^{d_2} u_2\/\ud y$. The horizontal velocities $u_j$ are thus (approximately) uniform along the layer column and the vertical velocities $v_j$ are chosen so that the fluid incompressibility are fulfilled together with the bottom and the lid impermeabilities.

With the ans\"atze (\ref{defuvsej}), the vertical accelerations are
\begin{align*}
  \uD_t\/v_j\ &\eqdef\ v_{jt}\ +\ u_j\,v_{jx}\ +\ v_j\,v_{jy}\ \approx\ \gamma_j\,h_j^{\,-1}\left\{d_j\,-\,(-1)^jy\right\},
\end{align*}
where $\uD_t$ is the temporal derivative following the motion and $\gamma_j$ are the vertical accelerations at the interface, \ie
\begin{align}\label{defgamma}
  \gamma_j\ &\eqdef\ \left.\uD_t\/v_j\right|_{y=\eta}\ \approx\ (-1)^j\,h_j\left\{\,\bar{u}_{jxt}\, +\, \bar{u}_j\,\bar{u}_{jxx}\, -\, \bar{u}_{jx}^{\,2}\,\right\}.
\end{align}
The kinetic and potential energies of the liquid column are, respectively, 
\begin{align*}
  \mathscr{K}\ &\eqdef\ \int_{-d_1}^\eta\rho_1\,\frac{u_1^{\,2}+v_1^{\,2}}{2}\,\ud\/y\ +\ \int_\eta^{d_2}\rho_2\,\frac{u_2^{\,2}+v_2^{\,2}}{2}\,\ud\/y\ \approx\ \sum\nolimits_{j=1}^2\rho_j\left(\,\half\,h_j\,\bar{u}_j^{\,2}\,+\,\sixth\,h_j^{\,3}\,\bar{u}_{jx}^{\,2}\,\right), \\
  \mathscr{V}\ &\eqdef\ \int_{-d_1}^\eta \rho_1\,g\left(y+d_1\right)\;\ud\/y\ +\ \int_\eta^{d_2} \rho_2\,g\left(y+d_1\right)\;\ud\/y\ =\ \half\,(\rho_1-\rho_2)\,g\,h_1^{\,2}\ +\ \half\,\rho_2\,g\,D^{2}.
\end{align*}

The incompressibility of the fluids and the impermeabilities of the lower and upper boundaries being fulfilled, a Lagrangian density $\mathscr{L}$ is then obtained from the Hamilton principle: the Lagrangian is the kinetic minus potential energies plus constraints for the mass conservation of each layer, \ie 
\begin{align*}
  \mathscr{L}\ &\eqdef\ \mathscr{K}\ -\  \mathscr{V}\ +\ \rho_1\left\{\,h_{1t}\, +\, \left[\,h_1\,\bar{u}_1\,\right]_x\,\right\}\phi_1\ +\ \rho_2\left\{\,h_{2t}\, +\, \left[\,h_2\,\bar{u}_2\,\right]_x\,\right\}\phi_2,
\end{align*}
where $\phi_j$ are \textsc{Lagrange} multipliers.

%%% ----------------------------------------------------------------------- %%%

\subsection{Equations of motion}

The \textsc{Euler}--\textsc{Lagrange} equations for the functional $\iint \mathscr{L}\,\ud\/x\,\ud\/t$ yield (together with $h_2 = D - h_1$ and for $j=1,2$)
\begin{align}
  \delta\phi_j:\ & 0\ =\ h_{jt}\ +\,\left[\,h_j\,\bar{u}_j\,\right]_x,\label{dLdphij}\\
  \delta\bar{u}_j:\ & 0\ =\ \phi_j\,h_{jx}\ -\ [\,h_j\,\phi_j\,]_x\ -\ \third\,[\,h_j^{\,3}\,\bar{u}_{jx}\,]_x\ +\ h_j\,\bar{u}_j, \label{dLduj}\\
  \delta h_1:\ & 0\ =\ \half\left(\,\rho_1\,\bar{u}_1^{\,2}\,-\,\rho_2\,\bar{u}_2^{\,2}\,\right)\ -\ (\rho_1-\rho_2)\,g\,h_1\ +\ \half\left(\,\rho_1\,h_1^{\,2}\,\bar{u}_{1x}^{\,2}\,-\,\rho_2\,h_2^{\,2}\,\bar{u}_{2x}^{\,2}\,\right)\nonumber\\ 
  &\qquad\ -\ \rho_1\,\phi_{1t}\ +\ \rho_2\,\phi_{2t}\ -\ \rho_1\,\bar{u}_1\,\phi_{1x}\ +\ \rho_2\,\bar{u}_2\, \phi_{2x}.\label{dLdh1}
\end{align}

Adding the two equations (\ref{dLdphij}) and integrating the result, one obtains
\begin{equation*}
  h_1\,\bar{u}_1\ +\ h_2\,\bar{u}_2\ =\ Q(t),
\end{equation*}
$Q$ being an integration `constant' ($U_\text{m}\eqdef Q/D$ is often called {\em mix velocity} in the theory of multiphase flows). The relation (\ref{dLduj})  can be rewritten
\begin{equation}\label{eqphi1x}
  \phi_{jx}\ =\ \bar{u}_j\ -\ \third\,h_j^{\,-1}\,[\,h_j^{\,3}\,\bar{u}_{jx}\,]_x\ =\ \bar{u}_j\ -\ \third\,h_j^{\,2}\,\bar{u}_{jxx}\ -\  h_j\,h_{jx}\,\bar{u}_{jx} \qquad (j = 1,2), 
\end{equation}
thence
\begin{align}
  h_1\,\phi_{1x}\ +\ h_2\,\phi_{2x}\ &=\ Q\ -\ \third\,[\,h_1^{\,3}\,\bar{u}_{1x}\,+\,h_2^{\,3}\,\bar{u}_{2x}\,]_x, \\
  \rho_1\,\phi_{1x}\ -\ \rho_2\,\phi_{2x}\ &=\ \rho_1\,\bar{u}_1\ -\ \rho_2\,\bar{u}_2\nonumber\\
  &\qquad -\ \third\,\rho_1\,h_1^{\,-1}\,[\,h_1^{\,3}\,\bar{u}_{1x}\,]_x\ +\ \third\,\rho_2\,h_2^{\,-1}\,[\,h_2^{\,3}\,\bar{u}_{2x}\,]_x, \label{phix12} \\
  \rho_1\,u_1\,\phi_{1x}\ -\ \rho_2\,u_2\,\phi_{2x}\ &=\ \rho_1\,\bar{u}_1^{\,2}\ -\ \rho_2\,\bar{u}_2^{\,2}\nonumber\\ 
  &\qquad-\ \third\,\rho_1\,u_1\,h_1^{\,-1}\,[\,h_1^{\,3}\,\bar{u}_{1x}\,]_x\ +\ \third\,\rho_2\,u_2\,h_2^{\,-1}\,[\,h_2^{\,3}\,\bar{u}_{2x}\,]_x. \label{uphix12} 
\end{align}
The equation (\ref{dLdh1}) then gives
\begin{align}\label{dLdh1bis}
  \rho_1\,\phi_{1t}\ -\ \rho_2\,\phi_{2t}\ &=\ \half\left(\,\rho_1\,h_1^{\,2}\,\bar{u}_{1x}^{\,2}\,-\,\rho_2\,h_2^{\,2}\,\bar{u}_{2x}^{\,2}\,\right)\,-\ \half\left(\,\rho_1\,\bar{u}_1^{\,2}\,-\,\rho_2\,\bar{u}_2^{\,2}\,\right)\ -\ (\rho_1-\rho_2)\,g\,h_1\nonumber\\ 
  &\quad +\ \third\,\rho_1\,\bar{u}_1\,h_1^{\,-1}\,[\,h_1^{\,3}\,\bar{u}_{1x}\,]_x\ -\ \third\,\rho_2\,\bar{u}_2\,h_2^{\,-1}\,[\,h_2^{\,3}\,\bar{u}_{2x}\,]_x\,,
\end{align}
and eliminating $\phi_j$ between (\ref{phix12}) and (\ref{dLdh1bis}), one obtains
\begin{align}
  &\partial_t\!\left\{\,\rho_1\,\bar{u}_1\, -\, \rho_2\,\bar{u}_2\, -\, \third\,\rho_1\,h_1^{\,-1}\,[\,h_1^{\,3}\,\bar{u}_{1x}\,]_x\, +\, \third\,\rho_2\,h_2^{\,-1}\,[\,h_2^{\,3}\,\bar{u}_{2x}\,]_x\,\right\}\,+ \nonumber \\
  &\partial_x\!\left\{\,\half\,\rho_1\,\bar{u}_1^{\,2}\,-\,\half\,\rho_2\,\bar{u}_2^{\,2}\, +\,(\rho_1-\rho_2)\,g\,h_1\,-\,\half\,\rho_1\,h_1^{\,2}\,\bar{u}_{1x}^{\,2}\,+\,\half\,\rho_2\,h_2^{\,2}\,\bar{u}_{2x}^{\,2}\right. \nonumber\\
  &\quad\ \left. -\,\third\,\rho_1\,\bar{u}_1\,h_1^{\,-1}\,[\,h_1^{\,3}\,\bar{u}_{1x}\,]_x\, +\, \third\,\rho_2\,\bar{u}_2\,h_2^{\,-1}\,[\,h_2^{\,3}\,\bar{u}_{2x}\,]_x\right\}\,=\ 0, \label{qdmcons}
\end{align}
that, physically, is an equation for the conservation of the difference between the tangential momenta at the interface. One can also easily derive a non-conservative equation for the horizontal momentum
\begin{align}
  &\rho_1\,(\bar{u}_{1t}\,+\,\bar{u}_1\,\bar{u}_{1x})\ -\ \rho_2\,(\bar{u}_{2t}\,+\,\bar{u}_2\,\bar{u}_{2x})\ +\ (\rho_1-\rho_2)\,g\,h_{1x} \nonumber\\
  &+\ \third\,\rho_1\,h_1^{\,-1}\,[\,h_1^{\,2}\,\gamma_1\,]_x\, +\ \third\,\rho_2\,h_2^{\,-1}\,[\,h_2^{\,2}\,\gamma_2\,]_x\ =\ 0.\label{qdm}
\end{align}
On the other hand, equations for the momentum and energy fluxes are not easily derived from these equations. This is where a multi-symplectic formulation comes to help.

%%% ----------------------------------------------------------------------- %%%

\section{Multi-symplectic structure}
\label{sec:ms}

A system of partial differential equations has a multi-symplectic structure if it can written as a system of first-order equations \cite{Bridges1997, Marsden1998}
\begin{equation}\label{sympgen}
  \mathbb{M}\scal\vz_t\ +\ \mathbb{K}\scal\vz_x\ =\ \grad_{\!z}\,S(\vz),
\end{equation}
where a dot denotes the contracted (inner) product, $\vz \in \mathds{R}^n$ is a rank-one tensor (vector) of state variables, $\mathbb{M} \in \mathds{R}^{n\times n}$ and $\mathbb{K} \in \mathds{R}^{n\times n}$ are skew-symmetric rank-two tensors (matrices) and $S$ is a smooth rank-zero tensor (scalar) function depending on $\vz$. $S$ is sometimes called the `\emph{Hamiltonian}', though it is generally not a classical Hamiltonian.

The multi-symplectic structure for the one-layer \textsc{Serre} equations is 
already known \cite{Chhay2016}. This structure can be easily extended to two (and more) layers. The multi-symplectic formulation for one layer involves $8$-by-$8$ matrices. For two layers, we then expect {\em a priori} a multi-symplectic formulation with $16$-by-$16$ matrices. However, since we consider a rigid lid for simplicity, one variable can be eliminated, thus reducing the formulation to $15$-by-$15$. Thus, introducing $h_1 = h$ and $h_2 = D - h$ for brevity, we seek for a multi-symplectic structure with
\begin{align}
  \vz\ &=\ h\,\ve_1\ +\ \varphi_1\,\ve_2\ +\ \bar{u}_1\,\ve_3\ +\ \tilde{v}_1\,\ve_4\ +\ p_1\,\ve_5\ +\ q_1\,\ve_6\ +\ r_1\,\ve_7\ +\ s_1\,\ve_8\nonumber\\
  &\quad\ +\ \varphi_2\,\ve_9\ +\ \bar{u}_2\,\ve_{10}\ +\ \tilde{v}_2\,\ve_{11}\ +\ p_2\,\ve_{12}\ +\ q_2\,\ve_{13}\ +\ r_2\,\ve_{14}\ +\ s_2\,\ve_{15}, \label{defz}
\end{align}
($\ve_i$ unitary standard basis vectors) and
\begin{align}
  \mathbb{M}\ &=\ \rho_1\left(\ve_1\otimes\ve_2\/-\/\ve_2\otimes\ve_1\right)\, +\ \third\,\rho_1\left(\ve_1\otimes\ve_5\/-\/\ve_5\otimes\ve_1\right)\nonumber\\
  &\quad\ -\ \rho_2\left(\ve_1\otimes\ve_9\/-\/\ve_9\otimes\ve_1\right)\, -\ \third\,\rho_2\left(\ve_1\otimes\ve_{12}\/-\/\ve_{12}\otimes\ve_1\right), \label{defM} \\
  \mathbb{K}\ & =\  \third\,\rho_1\left(\ve_1\otimes\ve_7\/-\/\ve_7\otimes\ve_1\right)\,-\ \rho_1\left(\ve_2\otimes\ve_6\/-\/\ve_6\otimes\ve_2\right)\nonumber\\
  &\quad\ -\ \third\,\rho_2\left(\ve_1\otimes\ve_{14}\/-\/\ve_{14}\otimes\ve_1\right)\,+\ \rho_2\left(\ve_9\otimes\ve_{13}\/-\/\ve_{13}\otimes\ve_9\right), \label{defK} \\
  S\ &=\ \rho_1\left(\/\sixth\/\tilde{v}_1^{\,2}\/-\/\half\/\bar{u}_1^{\,2}\/-\/\third\/s_1\/\bar{u}_1\/\tilde{v}_1\/\right)h\ +\ \rho_2\left(\/\sixth\/\tilde{v}_2^{\,2}\/-\/\half\/\bar{u}_2^{\,2}\/-\/\third\/s_2\/\bar{u}_2\/\tilde{v}_2\/\right)(D-h) \nonumber \\
  &\quad\ -\ \half\,(\rho_1-\rho_2)\,g\,h^2\ +\ \third\,\rho_1\,p_1\,(\/\bar{u}_1\/s_1\/-\/\tilde{v}_1\/)\ -\ \third\,\rho_2\,p_2\,(\/\bar{u}_2\/s_2\/-\/\tilde{v}_2\/)\nonumber \\ 
  &\quad\ +\ \rho_1\,q_1\/(\/\bar{u}_1\/+\/\third\/s_1\/\tilde{v}_1\/)\ -\ \rho_2\,q_2\/(\/\bar{u}_2\/+\/\third\/s_2\/\tilde{v}_2\/)\ -\ \third\,\rho_1\,r_1\,s_1\ +\ \third\,\rho_2\,r_2\,s_2. \label{defS}
\end{align}
These two-layer expressions for $\vz$, $\mathbb{M}$, $\mathbb{K}$ and $S$ are simple duplication of the corresponding expression for one layer \cite{Chhay2016}. We now show that they indeed lead to the two-layer \textsc{Serre}-like equations derived in the previous Section.

The substitution of (\ref{defz})--(\ref{defS}) into (\ref{sympgen}) yields the fifteen equations
\begin{gather}
%01
  \rho_1\left\{\,\varphi_{1t}\,+\,\third\,p_{1t}\,+\,\third\,r_{1x}\,\right\}\,-\ \rho_2\left\{\,\varphi_{2t}\,+\,\third\,p_{2t}\,+\,\third\,r_{2x}\,\right\}\,=\ -\,(\rho_1-\rho_2)\,g\,h \nonumber \\
  +\ \rho_1\left\{\,\sixth\,\tilde{v}_1^{\,2}\,-\,\half\,\bar{u}_1^{\,2}\,-\,\third\,\bar{u}_1\,\tilde{v}_1\,s_1\,\right\}\,-\,\rho_2\left\{\,\sixth\,\tilde{v}_2^{\,2}\,-\,\half\,\bar{u}_2^{\,2}\, - \,\third\,\bar{u}_2\,\tilde{v}_2\,s_2\,\right\}, \label{eqS01}\\
%02
  -\/\rho_1\,\{\,h_t\,+\,q_{1x}\,\}\ =\ 0, \label{eqS02}\\
%03
  0\ =\ \rho_1\/\left\{\,q_1\,-\,h\left(\,\bar{u}_1\,+\,\third\,\tilde{v}_1\,s_1\,\right)\, +\,\third\,p_1\,s_1\,\right\}, \label{eqS03}\\
%04
  0\ =\ -\/\third\,\rho_1\/\left\{\,p_1\,-\,h\left(\,\tilde{v}_1\,-\,\bar{u}_1\,s_1\,\right)\, -\,q_1\,s_1\,\right\}, \label{eqS04}\\
%05
  -\/\third\,\rho_1\,h_t\ =\ -\/\third\,\rho_1\left\{\,\tilde{v}_1\,-\,s_1\,\bar{u}_1\,\right\}, \label{eqS05}\\
%06
  \rho_1\,\varphi_{1x}\ =\ \rho_1\left\{\,\bar{u}_1\,+\,\third\,s_1\,\tilde{v}_1\,\right\}, \label{eqS06}\\
%07
  -\/\third\,\rho_1\,h_x\ =\ -\/\third\,\rho_1\,s_1, \label{eqS07}\\
%08
  0\ =\ -\/\third\,\rho_1\/\left\{\,r_1\,+\,h\,\bar{u}_1\,\tilde{v}_1\,-\,p_1\,\bar{u}_1\,-\,q_1\,\tilde{v}_1\,\right\}, \label{eqS08}\\
%09
  \rho_2\,\{\,h_t\,+\,q_{2x}\,\}\ =\ 0, \label{eqS09}\\
%10
  0\ =\ -\/\rho_2\/\left\{\,q_2\,+\,(D-h)\left(\,\bar{u}_2\,+\,\third\,\tilde{v}_2\,s_2\,\right)\, +\,\third\,p_2\,s_2\,\right\}, \label{eqS10}\\
%11
  0\ =\ \third\,\rho_2\/\left\{\,p_2\,+\,(D-h)\left(\,\tilde{v}_2\,-\,\bar{u}_2\,s_2\,\right)\, -\ q_2\,s_2\,\right\},  \label{eqS11}\\
%12
  \third\,\rho_2\,h_t\ =\ \third\,\rho_2\left\{\,\tilde{v}_2\,-\,s_2\,\bar{u}_2\,\right\},  \label{eqS12}\\
%13
  -\/\rho_2\,\varphi_{2x}\ =\ -\/\rho_2\left\{\,\bar{u}_2\,+\,\third\,s_2\,\tilde{v}_2\,\right\},  \label{eqS13}\\
%14
  \third\,\rho_2\,h_x\ =\ \third\,\rho_2\,s_2,  \label{eqS14}\\
%15
  0\ =\ \third\,\rho_2\/\left\{\,r_2\,-\,(D-h)\,\bar{u}_2\,\tilde{v}_2\,-\,p_2\,\bar{u}_2\,-\,q_2\,\tilde{v}_2\,\right\}.\label{eqS15}
\end{gather}
Twelve of these equations are trivial and can be simplified as (with $j=1,2$)
\begin{align*}
  q_j\,&=\,(-1)^{j-1}\,h_j\,\bar{u}_j, \qquad
  p_j\,=\,(-1)^{j-1}\,h_j\,\tilde{v}_j, \qquad s_j\, =\,(-1)^{j-1}\,h_{jx}, \nonumber\\
  r_j\,&=\, \bar{u}_j\,p_j, \qquad 
  \tilde{v}_j\, =\, (-1)^{j-1}\left(\,h_{jt}\, +\, \bar{u}_j\,h_{jx}\,\right), \qquad 
  \varphi_{jx}\, =\, \bar{u}_j\, +\, \third\,s_j\,\tilde{v}_j,
\end{align*}
the remaining three giving the mass conservation equations (together with $h_1+h_2=D$)
\begin{equation}\label{eqS13}
  h_{1t}\ +\,\left[\,h_1\,\bar{u}_1\,\right]_x\ =\ 0, \qquad 
  h_{2t}\ +\,\left[\,h_2\,\bar{u}_2\,\right]_x\ =\ 0,
\end{equation}
and, exploiting the relations (\ref{eqS13}), the equation for the tangential momenta at the interface
\begin{align}\label{eqS1bis}
  \rho_1\,\varphi_{1t}\ -\ \rho_2\,\varphi_{2t}\ &=\ \half\left(\,\rho_1\,h_1^{\,2}\,\bar{u}_{1x}^{\,2}\,-\,\rho_2\,h_2^{\,2}\,\bar{u}_{2x}^{\,2}\,\right)\,-\ \half\left(\,\rho_1\,\bar{u}_1^{\,2}\,-\,\rho_2\,\bar{u}_2^{\,2}\,\right)\nonumber\\ 
  &\quad +\ \third\,\rho_1\,\bar{u}_1\,h_1^{\,-1}\,[\,h_1^{\,3}\,\bar{u}_{1x}\,]_x\ -\ \third\,\rho_2\,\bar{u}_2\,h_2^{\,-1}\,[\,h_2^{\,3}\,\bar{u}_{2x}\,]_x \nonumber\\ 
  &\quad +\ \third\,\rho_1\,[\,h_1^{\,2}\,\bar{u}_{1x}\,]_t\ -\ \third\,\rho_2\,[\,h_2^{\,2}\,\bar{u}_{2x}\,]_t\ -\ (\rho_1-\rho_2)\,g\,h_1\,.
\end{align}
One can verify that these equations are equivalent to the ones obtained above.

The multi-symplectic structure described above involves the potentials $\varphi_j$ that are different from the potentials $\phi_j$ used to derive \textsc{Serre}-like equations of the section~\ref{sec:model}. After elimination of $\phi_j$ and $\varphi_j$, the two systems of equations are identical. Indeed, the differences between these two velocity potentials are --- from (\ref{eqphi1x}), (\ref{eqS12}) and (\ref{eqS13}) --- given by
\begin{align}
  \varphi_{jx}\ -\ \phi_{jx}\ &=\ \third\,h_{jx}\left(\,h_{jt}\,+\,u_j\,h_{jx}\,\right)\ +\ \third\,h_j^{\,2}\,\bar{u}_{jxx}\ +\  h_j\,h_{jx}\,\bar{u}_{jx} \nonumber\\
  &=\  \third\,h_j^{\,2}\,\bar{u}_{jxx}\ +\  \twothird\,h_j\,h_{jx}\,\bar{u}_{jx}\ =\ \third\left[\,h_j^{\,2}\,\bar{u}_{jx}\,\right]_x,
\end{align}
thence with \(\varphi_j=\phi_j\/+\/\third\/h_j^{\,2}\/\bar{u}_{jx}\) substituted into (\ref{eqS1bis}), the equation (\ref{dLdh1bis}) is recovered.

Physically, $\varphi_j$ are the velocity potentials written at the interface, while $\phi_j$ are related to the velocity potentials integrated over the fluid layers. Using the velocity potentials at the interface, we obtained rather easily the multi-symplectic structure of the \textsc{Serre}-like equations. Conversely, with the equivalent formulation involving $\phi_j$ it is difficult, and perhaps even impossible, to obtain a multi-symplectic structure of the \textsc{Serre}-like equations.

%%% ----------------------------------------------------------------------- %%%

\section{Conservation laws}
\label{sec:cons}

From the multi-symplectic structure, one obtains local conservation laws for the energy and the momentum
\begin{equation}
  E_t\ +\ F_x\ =\ 0, \qquad I_t\ +\ G_x\ =\ 0,
\end{equation}
where \(E(\vz)=S(\vz)+\half\/\vz_x\scal\mathbb{K}\scal\vz\), \(F(\vz)=-\half\/\vz_t\scal\mathbb{K}\scal\vz\), 
\(G(\vz)=S(\vz)+\half\/\vz_t\scal\mathbb{M}\scal\vz\) and \(I(\vz)=-\half\/\vz_x\scal\mathbb{M}\scal\vz\).
For the \textsc{Serre}-like equations, exploiting the results of the previous section and after some algebra, one obtains
\begin{align}
  E\ &=\ \rho_1\left[\,\half\,\varphi_1\,h_1\,\bar{u}_1\,+\,\sixth\,h_1^{\,3}\,\bar{u}_1\,\bar{u}_{1x}\,\right]_x\ -\,\half\,\rho_1\,h_1\,\bar{u}_1^{\,2}\ -\ \sixth\,\rho_1\,h_1^{\,3}\,\bar{u}_{1x}^{\,2} \nonumber\\
  &\quad\, +\ \rho_2\left[\,\half\,\varphi_2\,h_2\,\bar{u}_2\,+\,\sixth\,h_2^{\,3}\,\bar{u}_2\,\bar{u}_{2x}\,\right]_x\ -\ \half\,\rho_2\,h_2\,\bar{u}_2^{\,2}\ -\ \sixth\,\rho_2\,h_2^{\,3}\,\bar{u}_{2x}^{\,2}\nonumber\\
  &\quad\, -\ \sixth\,\rho_2\,D\left[\,h_2^{\,2}\,\bar{u}_2\,\bar{u}_{2x}\,\right]_x\ -\ \half\,(\rho_1-\rho_2)\,g\,h_1^{\,2}, \label{expE}\\
  F\ &=\, -\,\rho_1\left[\,\half\,\varphi_1\,h_1\,\bar{u}_1\,+\,\sixth\,h_1^{\,3}\,\bar{u}_1\,\bar{u}_{1x}\,\right]_t\ -\ \rho_1\,h_1\,\bar{u}_1\left(\,\half\,\bar{u}_1^{\,2}\,+\,\sixth\,h_1^{\,2}\,\bar{u}_{1x}^{\,2}\,+\,\third\,\,h_1\,\gamma_1\,\right) \nonumber\\
  &\quad\ -\ \rho_2\left[\,\half\,\varphi_2\,h_2\,\bar{u}_2\,+\,\sixth\,h_2^{\,3}\,\bar{u}_2\,\bar{u}_{2x}\,\right]_t\ -\ \rho_2\,h_2\,\bar{u}_2\left(\,\half\,\bar{u}_2^{\,2}\,+\,\sixth\,h_2^{\,2}\,\bar{u}_{2x}^{\,2}\,-\,\third\,\,h_2\,\gamma_2\,\right)\nonumber\\ 
  &\quad\ +\ \sixth\,\rho_2\,D\left[\,h_2^{\,2}\,\bar{u}_2\,\bar{u}_{2x}\,\right]_t\ -\ (\rho_1-\rho_2)\,g\,h_1^{\,2}\,\bar{u}_1\ +\ \rho_2\,Q\,\phi_{2t}\ +\ \half\,\rho_2\,Q\,\bar{u}_2^{\,2} \nonumber\\
  &\quad\ -\ \rho_2\,Q\left[\,\half\,h_2^{\,2}\,\bar{u}_2\,\bar{u}_{2x}\,\right]_x\ +\ \sixth\,\rho_2\,Q\,h_2^{\,2}\,\bar{u}_2\,\bar{u}_{2xx}, \label{expF} \\
  G\ &=\ \rho_1\,h_1\,\bar{u}_1^{\,2}\ +\ \half\,(\rho_1-\rho_2)\,g\,h_1^{\,2}\ +\ \third\,\rho_1\,h_1^{\,2}\,\gamma_1\ +\ \rho_1\left[\,\half\,\varphi_1\,h_1\,+\,\sixth\,h_1^{\,3}\,\bar{u}_{1x}\,\right]_t \nonumber\\
  &\quad -\ \half\,\rho_2\,(h_1-h_2)\,\bar{u}_2^{\,2}\ -\ \third\,\rho_2\,h_2^{\,2}\,\gamma_2\ +\ \rho_2\left[\,\half\,\varphi_2\,h_2\,+\,\sixth\,h_2^{\,3}\,\bar{u}_{2x}\,\right]_t \nonumber\\
  &\quad -\ \half\,\rho_2\,D\,\phi_{2t}\ +\ \half\,\rho_2\,D\left[\,h_2^{\,2}\,\bar{u}_2\,\bar{u}_{2x}\,\right]_x\ -\ \sixth\,\rho_2\,D\,h_2^{\,2}\,\bar{u}_2\,\bar{u}_{2xx}, \label{expG}
\end{align}
\begin{align}
  I\ &=\ \rho_1\,h_1\,\bar{u}_1\ -\ \rho_1\left[\,\half\,\varphi_1\,h_1\,+\,\sixth\,h_1^{\,3}\,\bar{u}_{1x}\,\right]_x\ -\  \half\,\rho_2\,D\,\phi_{2x} \nonumber\\ 
  &\quad +\ \rho_2\,h_2\,\bar{u}_2\ -\ \rho_2\left[\,\half\,\varphi_2\,h_2\,+\,\sixth\,h_2^{\,3}\,\bar{u}_{2x}\,\right]_x. \label{expI}
\end{align}
Note that these relations involve both $\varphi_j$ and $\phi_j$ in order to handle more compact expressions.

%%% ----------------------------------------------------------------------- %%%

\subsection{Momentum flux}

From the relations (\ref{expG}) and (\ref{expI}), after simplifications and some elementary algebra, one obtains the equation for the conservation of the momentum flux
\begin{gather}
  \sum\nolimits_{j=1}^2\ \partial_t\!\left[\,\rho_j\,h_j\,\bar{u}_j\,\right]\ +\ \partial_x\!\left[\,\rho_j\,h_j\left\{\,\bar{u}_j^{\,2}\, -\, (-1)^{j}\,\half\,g\,h_j\,-\,(-1)^{j}\,\third\,h_j\,\gamma_j\,\right\}\right]\,= \nonumber\\
  D\,\rho_2\,\partial_x\!\left[\,\phi_{2t}\ +\ \half\,\bar{u}_2^{\,2}\ -\ h_2\,h_{2x}\,\bar{u}_2\,\bar{u}_{2x}\ -\ \half\,h_2^{\,2}\,\bar{u}_{2x}^{\,2}\ -\ \third\,h_2^{\,2}\,\bar{u}_2\,\bar{u}_{2xx}\ -\ g\,h_2\,\right],\label{DPix0}
\end{gather}
and comparison with the equation (\ref{DPix}) gives an expression for the pressure at the interface
\begin{equation}\label{Pintupp}
  \tilde{P}\,/\,\rho_2\ =\ K_2(t)\ -\ \phi_{2t}\ -\ \half\,\bar{u}_2^{\,2}\ +\ h_2\,h_{2x}\,\bar{u}_2\,\bar{u}_{2x}\ +\ \half\,h_2^{\,2}\,\bar{u}_{2x}^{\,2}\ +\ \third\,h_2^{\,2}\,\bar{u}_2\,\bar{u}_{2xx}\ +\ g\,h_2\,,
\end{equation}
where $K_2(t)$ is an arbitrary function (integration `constant'). The horizontal derivative of this relation, with (\ref{dLdphij}) and (\ref{eqphi1x}), yields after some algebra
\begin{equation*}
  \tilde{P}_x\,/\,\rho_2\ =\ g\,h_{2x}\ -\ \bar{u}_{2t}\ -\ \bar{u}_2\,\bar{u}_{2x}\ +\ \third\,h_2^{\,-1}\left[\,h_2^{\,2}\,\gamma_2\,\right]_x,
\end{equation*}
that is the upper-layer averaged horizontal momentum equation (\ref{Pixj}), as it should be. We have obtained the \textsc{Cauchy}--\textsc{Lagrange} equation (\ref{Pintupp}) because we used $h\ =\ h_1$ as main variable for the interface in the multi-symplectic formalism and because we eliminated $\phi_1$ from the equations. Had we instead used $h_2$ and eliminated $\phi_2$, we would have obtained a \textsc{Cauchy}--\textsc{Lagrange} equation for the lower layer. The latter can be easily derived in the form   
\begin{equation*}
  \tilde{P}\,/\,\rho_1\ =\ K_1(t)\ -\ \phi_{1t}\ -\ \half\,\bar{u}_1^{\,2}\ +\ h_1\,h_{1x}\,\bar{u}_1\,\bar{u}_{1x}\ +\ \half\,h_1^{\,2}\,\bar{u}_{1x}^{\,2}\ +\ \third\,h_1^{\,2}\,\bar{u}_1\,\bar{u}_{1xx}\ -\ g\,h_1.
\end{equation*}

%%% ----------------------------------------------------------------------- %%%

\subsection{Energy flux}

From the relations (\ref{expE}) and (\ref{expF}), after simplifications and some elementary algebra, one obtains the equation for the conservation of the energy flux
\begin{gather}
  \sum\nolimits_{j=1}^2\ \partial_t\!\left[\,\half\,\rho_j\,h_j\left\{\,\bar{u}_j^{\,2}\,+\,\third\,h_j^{\,2}\,\bar{u}_{jx}^{\,2}\,-\,(-1)^{j}\,g\,h_j\,\right\}\rule{0mm}{4mm}\right]\ +\nonumber\\
  \partial_x\!\left[\,\rho_j\,h_j\,\bar{u}_j\left\{\,\half\,\bar{u}_j^{\,2}\,+\,\sixth\,h_j^{\,2}\,\bar{u}_{jx}^{\,2}\,-\,(-1)^{j}\,\third\,h_j\,\gamma_j\,-\,(-1)^{j}\,g\,h_j\,\right\}\rule{0mm}{4mm}\right]\,= \nonumber\\
  Q\,\rho_2\,\partial_x\!\left[\,\phi_{2t}\ +\ \half\,\bar{u}_2^{\,2}\ -\ h_2\,h_{2x}\,\bar{u}_2\,\bar{u}_{2x}\ -\ \half\,h_2^{\,2}\,\bar{u}_{2x}^{\,2}\ -\ \third\,h_2^{\,2}\,\bar{u}_2\,\bar{u}_{2xx}\ -\ g\,h_2\,\right].\label{DPixene}
\end{gather}
The right-hand sides of equations (\ref{DPix0}) and (\ref{DPixene}) both involve $\phi_2$. The latter can be eliminated computing $D\times(\ref{DPixene})-Q\times(\ref{DPix0})$, yielding after one integration by parts 
\begin{align}
  \sum\nolimits_{j=1}^2\ \partial_t\!\left[\,\half\,D\,\rho_j\,h_j\left\{\,\bar{u}_j^{\,2}\,+\,\third\,h_j^{\,2}\,\bar{u}_{jx}^{\,2}\,-\,(-1)^{j}\,g\,h_j\,\right\}\,-\ Q\,\rho_j\,h_j\,\bar{u}_j\,\rule{0mm}{4mm}\right]&\nonumber\\
  +\ \partial_x\!\left[\,D\,\rho_j\,h_j\,\bar{u}_j\left\{\,\half\,\bar{u}_j^{\,2}\,+\,\sixth\,h_j^{\,2}\,\bar{u}_{jx}^{\,2}\,-\,(-1)^{j}\,\third\,h_j\,\gamma_j\,-\,(-1)^{j}\,g\,h_j\,\right\}\rule{0mm}{4mm}\right.\ \, & \nonumber\\
  - \left.\,Q\,\rho_j\,h_j\left\{\,\bar{u}_j^{\,2}\, -\, (-1)^{j}\,\half\,g\,h_j\,-\,(-1)^{j}\,\third\,h_j\,\gamma_j\,\right\}\rule{0mm}{4mm}\right]&\nonumber \\
  =\ -\,\frac{\ud\,Q}{\ud\/t}\left\{\,\rho_1\,h_1\,\bar{u}_1\,+\,\rho_2\,h_2\,\bar{u}_2\,\right\} =\ -\,\rho_2\,Q\,\frac{\ud\,Q}{\ud\/t}\, -\ \frac{\ud\,Q}{\ud\/t}\,(\rho_1-\rho_2)\,h_1\,\bar{u}_1. &\label{DPixeneqdm}
\end{align}

%%% ----------------------------------------------------------------------- %%%

\section{Conclusions and perspectives}
\label{sec:concl}

We have derived the fully-nonlinear and weakly-dispersive \textsc{Serre}-type equations from a variational framework. The main contribution of this study is that we present their multi-symplectic structure. The rather complicated nonlinear dispersion of the \textsc{Serre}-like equations makes non-trivial the derivation of the multi-symplectic structure, {\em a priori}. However, we have shown here that this structure for fluids stratified in several homogeneous layers can be easily obtained from the one layer case \cite{Chhay2016}. For the sake of simplicity, we focused here on two layers in two dimensions with horizontal bottom and lid. Generalisations in three dimension and several layers are straightforward.

The size of the multi-symplectic structure increases rapidly with the number of layers and the number of spatial dimensions. However, the matrices involved are sparse and most of the equations are algebraically elementary. Thus, the calculus can be achieved easily and straightforwardly with any Computer Algebra System capable of performing symbolic computations.

Finding a multi-symplectic structure opens up new directions in the analysis and numerics of equations. In this paper, we illustrate another advantage of the multi-symplectic formalism: it also provides an efficient tool of calculus. Thanks to the multi-symplectic formalism, the conservation laws are obtained automatically, and thus the conserved quantities and their fluxes are obtained as well. The `automatic' derivation of these conservation laws is an advantage of the multi-symplectic structure.

As the main perspective, we would like to mention the structure-preserving numerical simulations of nonlinear internal waves. The proposed multi-symplectic structure can be transposed to the discrete level if one employs a multi-symplectic integrator \cite{Bridges2001, Moore2003a}. These schemes were already rested in complex KdV simulations \cite{Dutykh2013a}, but this direction seems to be essentially open for internal wave models.

%%% ----------------------------------------------------------------------- %%%

\appendix
\section{Complementary equations}
\label{sec:app}

Multiplying by $(y\ +\ d_1)$ and  $(d_2\ -\ y)$ the vertical momentum (full \textsc{Euler}) equation for the lower and upper layers, respectively, and integrating over the layer thicknesses, we have
\begin{align*}
  \int_{-d_1}^\eta\rho_1\,(y+d_1)\left[\,\frac{\uD\,v_1}{\uD\/t}\,+\,g\,\right]\ud\/y\ &=\ -\int_{-d_1}^\eta (y+d_1)\,\frac{\partial\,P_1}{\partial\/y}\,\ud\/y\ =\ h_1\left(\bar{P}_1\ -\ \tilde{P}\right), \\
  \int^{d_2}_\eta\rho_2\,(d_2-y)\left[\,\frac{\uD\,v_2}{\uD\/t}\,+\,g\,\right]\ud\/y\ &=\ - \int^{d_2}_\eta (d_2-y)\,\frac{\partial\,P_2}{\partial\/y}\,\ud\/y\ =\ h_2\left(\tilde{P}\ -\ \bar{P}_2\right),
\end{align*}
where $\tilde{P}$ is the (unknown) pressure at the interface and $\bar{P}_j$ is the layer-averaged pressure. With the ans\"atze (\ref{defuvsej}) one obtains
\begin{align*}
  \bar{P}_j\ &=\ \tilde{P}\ +\ (-1)^{\,j-1}\,\half\,\rho_j\,g\,h_j\ +\ (-1)^{\,j-1}\,\third\,\rho_j\,h_j\,\gamma_j, \qquad (j\ =\ 1,\,2).
\end{align*}
Integrating over the layer thicknesses the horizontal momenta, one obtains 
\begin{align*}
  \int_{-d_1}^\eta\rho_1\,\frac{\uD\,u_1}{\uD\/t}\;\ud\/y\ &=\ - \int_{-d_1}^\eta \frac{\partial\,P_1}{\partial\/x}\,\ud\/y\ =\, \left[\,h_1\left(\tilde{P}-\bar{P}_1\right)\right]_x\ -\ h_1\,\tilde{P}_x, \\
  \int^{d_2}_\eta\rho_2\,\frac{\uD\,u_2}{\uD\/t}\;\ud\/y\ &=\ -\int^{d_2}_\eta \frac{\partial\,P_2}{\partial\/x}\,\ud\/y\ =\, \left[\,h_2\left(\tilde{P}-\bar{P}_2\right)\right]_x\ -\ h_2\,\tilde{P}_x,
\end{align*}
thence with the ans\"atze (\ref{defuvsej}) (with $j=1,2$)
\begin{align*}
  \left[\,h_j\,\bar{u}_j\,\right]_t\ +\ \left[\,h_j\,\bar{u}_j^{\,2}\ -\ (-1)^j\,\half\,g\,h_j^{\,2}\ -\ (-1)^j\,\third\,h_j^{\,2}\,\gamma_j\,\right]_x\ &=\ -\,\rho_j^{\,-1}\,h_j\,\tilde{P}_x. 
\end{align*}
From these relations, we obtain at once (with $\dot{Q}=\ud Q/\ud t$)
\begin{gather}
  \dot{Q}\ +\ \left[\,h_1\,\bar{u}_1^{\,2}\,+\,h_2\,\bar{u}_2^{\,2}\,+\,g\,D\,h_1\, + \,\,\third\,h_1^{\,2}\,\gamma_1\,-\,\third\,h_2^{\,2}\,\gamma_2\,\right]_x\ =\ -\left(h_1\,\rho_1^{\,-1}\,+\,h_2\,\rho_2^{\,-1}\right)\tilde{P}_x, \label{hPix}\\
  \sum\nolimits_{j=1}^2\rho_j\left\{\,\left[\,h_j\,\bar{u}_j\,\right]_t\ +\, \left[\,h_j\,\bar{u}_j^{\,2}\,-\,(-1)^{j}\,\half\,g\,h_j^{\,2}\,-\,(-1)^{j}\,\third\,h_j^{\,2}\,\gamma_j\,\right]_x\,\right\}\, =\ -\,D\,\tilde{P}_x, \label{DPix}\\
  \bar{u}_{jt}\ +\ \bar{u}_j\,\bar{u}_{jx}\ -\ (-1)^j\,g\,h_{jx}\ -\ (-1)^j\,\third\,h_j^{\,-1}\left[\,h_j^{\,2}\,\gamma_j\,\right]_x\ =\ -\,\rho_j^{\,-1}\,\tilde{P}_x. \label{Pixj}
\end{gather}
The elimination of $\tilde{P}_x$ between the two equations (\ref{Pixj}) yields (\ref{qdm}).

%%% ----------------------------------------------------------------------- %%%

\subsection*{Acknowledgments}
\addcontentsline{toc}{subsection}{Acknowledgments}

The authors would like to acknowledge the support from CNRS under the PEPS 2015 Inphyniti programme and exploratory project \textsf{FARA}.

%%% ----------------------------------------------------------------------- %%%

%%% Bibliography
\addcontentsline{toc}{section}{References}
\bibliographystyle{abbrv}
\bibliography{biblio}

\end{document}